%% file: main.tex
\documentclass[sigconf,natbib=true]{acmart}

\AtBeginDocument{%
  \providecommand\BibTeX{{%
    \normalfont B\kern-0.5em{\scshape i\kern-0.25em b}\kern-0.8em\TeX}}}
\usepackage[title]{appendix}
\usepackage{dirtytalk}
\usepackage{subcaption}
\usepackage{bbm}
\usepackage[export]{adjustbox}
\usepackage{multirow}
\usepackage{makecell}
\usepackage[utf8]{inputenc}
\usepackage[T1]{fontenc}
\usepackage{tikz}
\usetikzlibrary{arrows,calc,positioning}
\usepackage{listings}
\usepackage{pbox}
\usepackage{amsmath} 
\usepackage{appendix}

\setcopyright{acmcopyright}
\copyrightyear{2018}
\acmYear{2018}
\acmDOI{XXXXXXX.XXXXXXX}

\acmConference[Conference acronym 'XX]{Make sure to enter the correct
  conference title from your rights confirmation emai}{June 03--05,
  2018}{Woodstock, NY}
\acmPrice{15.00}
\acmISBN{978-1-4503-XXXX-X/18/06}

\begin{document}


\title{The Impact of Snippet Reliability on Misinformation in Online Health Search
}


\author{Anat Hashavit$^1$, \  Tamar Stern$^1$, \ Hongning Wang$^2$, \  Sarit Kraus$^1$}
\affiliation{%
  \institution{$^1$Bar Ilan University,  $^2$University of Virginia}
  \country{$^1$Ramat~Gan, Israel, $^2$Charlottesville, VA, USA}}
\email{{anat.hashavit,stern.tamar96}@gmail.com, sarit@cs.biu.ac.il, hw5x@virginia.edu}

\renewcommand{\shortauthors}{Hashavit et al.}


\begin{abstract}
Search result snippets play a vital role in modern search engines by offering users a quick preview of a website's content. Snippets can assist users to quickly determine if a document is pertinent to their information need, and in certain scenarios even enable 
them to satisfy their information need without visiting the web document. 
Hence, it is crucial for snippets to reliably represent the content of their corresponding documents.
While this may be a straightforward requirement for simple factual queries, it becomes considerably more challenging in the complex domain of healthcare, and can lead to misinformation.

This paper aims to examine snippets' reliability in representing  their corresponding documents,
specifically in the health domain. To achieve this, we conduct a series of user studies using Google's search results, where participants are asked to infer the viewpoints of search results pertaining to queries about the effectiveness of a medical intervention for a specific medical condition, based solely on their titles and snippets. 
Our findings reveal that a considerable portion of Google's snippets (28\%) failed to present any viewpoint on the intervention's effectiveness, and an additional 35\% were interpreted by participants as having a different viewpoint compared to their corresponding web documents.
To address this issue, we propose a snippet extraction solution tailored directly to users' information needs, i.e., extracting snippets that summarize documents' viewpoints regarding the medical intervention and condition that appear in the query.   
User study demonstrates that our information need-focused solution outperforms the mainstream query-based approach used by commercial search engines. 
With only 19.67\% of snippets generated by our solution reported as not presenting a viewpoint and a mere 20.33\% being misinterpreted by participants.
 These results strongly
suggest that an information need-focused approach can significantly improve the reliability of extracted snippets in online health search.

\end{abstract}



\maketitle

\input{intro}
\input{relwork}

\input{user_study}

\input{google_perf.tex}
\input{algorithm}

\section{Discussion and conclusions}
\label{sec:conc}

In this work, we explored the reliability of snippets extracted for search results in the health domain, specifically relating to medical intervention effectiveness. We conducted a set of user studies in which participants were required to annotate the viewpoints of documents based on their titles and snippets alone. We compared snippets extracted by a commercial search engine to snippets manually extracted by a human reader, and saw a significant gap between the two methods' results. We showed that oftentimes the snippets extracted by the commercial search engine did not present the viewpoint of the document. We additionally showed that even when snippets did present a viewpoint, this viewpoint could often differ from that of the document.

We then proposed a light-weight information need-focused framework that extracted snippets by using a supervised learning model that classifies documents' viewpoints. 
We showed that even this simple approach can significantly outperform the query-based method for all forms of documents' viewpoints.

In future work we would like to further explore this phenomenon's scale in commercial search engines, by inspecting a wider variety of conditions and interventions, as well as a wider variety of query phrasing forms. We also plan to further explore the technical aspects of the information need-focused snippet extraction problem. In particular, we would like to construct a 
model that can directly learn the snippet extraction task in relation to intervention and condition terms that it will receive as input. 


\bibliographystyle{ACM-Reference-Format}
\bibliography{main}

\end{document}

%% file: intro.tex
\section{Introduction}
\begin{figure}[t]
\centering
\includegraphics[scale=0.5,keepaspectratio]{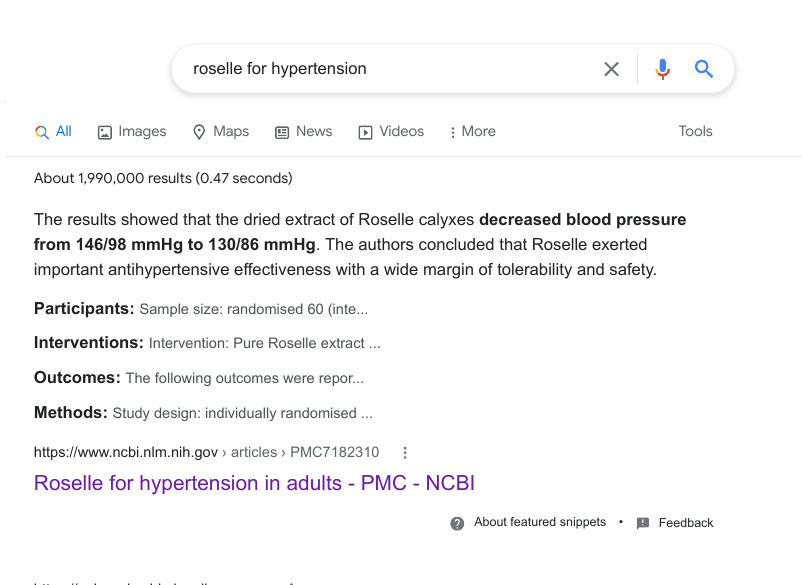}
\vspace{-7mm}
\caption{Results for the query: ``roselle for hypertension''. }
\label{fig:example}
\vspace{-7mm}
\end{figure}

Search engines mediate the interactions between information seekers and information providers. 
To help users quickly determine if a returned result is relevant to their information need, it is a common practice for search engines to describe each returned result by its title and a short snippet extracted by some algorithmic approach. In most cases, the snippet is extracted from the document based on the matching against the entered query, i.e., query-based \cite{GoogleSnippets,tombros1998advantages}. 

In the early days of search engines, snippets were simply short text excerpts extracted from web pages based on the search query \cite{tombros1998advantages}. These snippets were often limited to displaying the meta description or a portion of the page's content that matched the query.
As search engine techniques have advanced, so has the quality of the presented snippets. 
Notable examples of this progress include rich snippets, which present additional structured data from the web page \cite{goel2009introducing,hop2010automatic}, and featured snippets, designed to provide quick answers to users' queries within the SERP \cite{oliveira2023evolution,FeaturedSnippets}. 
These advancements aim to provide users with prompt answers and reduce the need for additional clicks, thereby improving overall search efficiency. 

As a result, in recent years an increasing number of users have been able to satisfy their information need without visiting any of the returned URLs \cite{li2009good, diriye2012leaving, stamou2010interpreting}. 
Given this phenomenon, it is crucial for snippets to represent their corresponding documents
in such a way that when a user's information need is met by reading a snippet, they will arrive at the same conclusion as if they had read the entire document. 
This requirement may appear trivial for straightforward factual queries where the vast majority of documents present the same response. However, when the answer to a query is not a clear-cut unanimous response, 
the wrong snippet can lead to the spread of misinformation.


Figure \ref{fig:example} presents a portion of the SERP for the query ``Roselle for hypertension'' from Google.
The first result is a very relevant document that details a scientific review with the exact same title as the query phrase. It is presented along with a featured snippet.
When reading the snippet, one can notice the bold portion of the text: \textit{``decreased blood pressure from 146/98 mmHg to 130/86 mmHg''}. Further reading of the snippet will also reveal the sentence: \textit{``The authors concluded that Roselle exerted important antihypertensive effectiveness with a wide margin of tolerability and safety.''}
If one visits this website, he/she will find a long scientific paper describing a literature review of research papers discussing the effectiveness of the Roselle plant in treating hypertension \cite{pattanittum2021roselle}. 
The presented snippet describes one of the papers included in the review, the conclusion of which is unfortunately quite different from the review article's conclusion. The conclusion section of the retrieved article reveals the following text:
\textit{``The available evidence regarding the effectiveness and safety of Roselle for hypertensive adults compared to placebo or no intervention is very uncertain and did not permit the drawing of any conclusions''}.
A user satisfied by only reading the snippet 
may mistakenly believe that Roselle is an effective treatment for hypertension, while the actual conclusion of the web document suggests that its effectiveness remains unclear. 

We refer to a snippet that accurately reflects the viewpoint of the document from which it is derived as a reliable snippet.
In this work, we investigate snippets' reliability in online health search. As in previous work on misinformation in the healthcare domain, we too focus on queries  
where the underlying  information need of the user is to determine the effectiveness of a medical intervention in treating a specific medical condition \cite{white2013beliefs, white2014content, ghenai2020think, pogacar2017positive, hashavit2021understanding}. 
Queries related to this information need hold significant importance because, unlike definitional or factual queries, the answers to these inquiries are often not unanimous: even a single document can present different viewpoints, as exemplified by the review article shown in Figure  ~\ref{fig:example}.
Therefore, it is crucial that when users rely on a snippet to meet such information needs, the snippet faithfully reflects the viewpoint of its corresponding document regarding the intervention's effectiveness in treating the condition. Consequently, addressing inaccuracies in snippets for such information needs becomes vital, as these inaccuracies have the potential to propagate misinformation, posing risks to the well-being of users.


Our work is divided into two parts. In the first part,
we examine the reliability of query-based snippets in the commonly used Google search engine \cite{GoogleSnippets}. 
We conduct a series of user studies in which participants are asked to evaluate the viewpoints of web documents regarding a medical intervention's effectiveness for a given condition, solely based on their snippets and titles.
We compare participants' performance using snippets extracted by Google's query-based method to that of a gold standard set, which consists of snippets manually extracted by a trained human reader.

Our observations reveal that when presented with snippets extracted by Google, participants frequently report that the viewpoint of the document was not presented in the snippet. Moreover, even when participants believe they can identify the viewpoint within a document's snippet, it often does not align with the viewpoint expressed in the document itself. Conversely, the gold standard snippets consistently yield significantly better results across all~metrics. 


We posit that this phenomenon stems from an inherent discrepancy between the query-based snippet extraction method and the information need expressed in the query phrases. In our study, the specified information need is to ascertain the effectiveness of an intervention in treating a specific health condition. Therefore, a reliable snippet should accurately summarize the viewpoint of its corresponding document regarding the effectiveness of the intervention.
However, the query-based snippet extraction approach is not designed to comprehend the document itself; it simply aims to identify the portion of the document that aligns most closely with the user's query phrase.

Consequently, the second part of our work delves into an information need-focused approach for snippet extraction, particularly for queries centered around the effectiveness of medical interventions. Our proposed solution incorporates two key components: a supervised learning-based algorithm for viewpoint detection, and a snippet extraction component that 
relies on the viewpoint classifier.
To extract information need-focused snippets, our solution leverages an explainable text classification method in which we select, as a snippet, the sentence that contributes the most to the document's viewpoint classification.
This approach significantly simplifies the labeling process by requiring document-level annotations instead of sentence-level labeling, resulting in a lightweight yet highly~effective solution.

To evaluate our solution, we utilize  the same user study protocol while employing snippets that are extracted using our proposed approach.  The results demonstrate that the information need-focused approach outperforms the query-based approach significantly. Users exhibit enhanced capabilities in identifying a viewpoint within a snippet and accurately inferring the viewpoint of the corresponding document when employing our solution.


%% file: relwork.tex
\vspace{-4mm}
\section{Related Work}
\label{sec:prev_work}

The research relating to search result snippets is mainly focused on two fields. The first is from the cognitive psychological aspect, researched within the human computer interaction (HCI) community. The second is in the computational and algorithmic research field. Our work touches on both fields, therefore we will discuss the related work in each field accordingly.

Early HCI research focused on the effects of snippets on users' search experience in general.
Tombros and Sanderson  \cite{tombros1998advantages} compared query-based snippets to a query-agnostic summary composed by the document's first few sentences. They showed that query-based snippets improve users' search experience by helping them satisfy their information need more quickly and evaluate documents' relevance more accurately. Indeed, a query-based snippet extraction method was the chosen approach in the Google search engine \cite{brin1998anatomy,GoogleSnippets}.
White et al. \cite{white2003task} conducted a task-oriented study on the influence of query-based snippets in web search. They also found that query-based snippets better help users assess document relevance than abstract-based summaries.
Since then several studies have explored various aspects of snippets and repeatedly found them to have a significant effect both on users' experience and their performance in their information seeking tasks
\cite{clarke2007influence,rose2007summary,maxwell2017study,cutrell2007you}.

Several studies explored a phenomenon called good abandonment, where users satisfy their information need without visiting any of the hyperlinks presented in the SERP,
  and they found this phenomenon to be quite prevalent \cite{li2009good, diriye2012leaving, stamou2010interpreting}. Diriye et al.  found that about a third of SERP abandonments are due to users having their information need satisfied by the SERP \cite{diriye2012leaving}. Stamou and Efthimiadis  \cite{stamou2010interpreting} further found that, in some cases, users even declared that their expectation was to satisfy their information need by relying only on the information presented in the SERP. This includes direct answers, featured snippets and regular snippets. 

The credibility of the information presented in the SERP has also been explored.
Wu et al. \cite{wu2020credibility} offer a model to evaluate how users perceive credibility of good abandonment results in mobile searches. They identified six factors with which users evaluate the quality of good abandonment results. Bink et al. 
\cite{bink2023investigating} explored the effect of featured snippets on users' attitudes towards debatable topics.
The study showed that users can change their attitudes and beliefs based on the content they consume in a featured snippet, without visiting its corresponding document. 

In another study conducted by Bink et al. \cite{bink2022featured}, the authors explored the impact of featured snippets on users' beliefs regarding the effectiveness of medical treatments. Their findings revealed that users often overestimate the effectiveness of medical interventions when relying solely on featured snippets. Consequently, depending on featured snippets to satisfy users' information needs can mislead them. However, the study did not investigate whether or not the snippets accurately represent their corresponding documents. 

The findings of the aforementioned study are consistent with several previous studies that have explored the issue of misinformation in online health searches from various perspectives 
~\cite{white2013beliefs, white2014content, ghenai2020think, pogacar2017positive, hashavit2021understanding, hashavit2022not}. These studies consistently found that users can be misled when seeking health-related information online. 
Specifically, they revealed that users are vulnerable to positive biases introduced by search engines and, as a result, can be convinced of the effectiveness of unproven medical interventions.
All of these studies focused on misinformation introduced by inaccuracies and biases present in the selected documents themselves. In this study, we bring forth another potential source of misinformation, where the document itself is accurate but the snippet representing it does not accurately reflect its underlying conclusion, thereby introducing bias to the search results \emph{even when the selected document itself does not}.


The algorithmic research field related to snippet extraction focuses mainly on developing tools to generate high quality snippets. 
Some algorithms use machine learning to enhance snippets' readability \cite{kanungo2009predicting}, while others use IR tools such as query expansion \cite{leal2015query}. Personalized snippet generation has also been studied by incorporating language models \cite{li2010personalized,ageev2013improving} 
Our work also proposes a snippet extraction solution, with a focus on online health search. The goal of our extraction method is to isolate a specific portion of the text that best corresponds to  the user’s core information need within the document, thereby effectively mitigating the risks of misinformation.

An information need-focused approach for argument mining \cite{stede2018argumentation,lippi2016argumentation} was discussed by Alshomary et al. \cite{alshomary2020extractive}.
The authors used argument mining techniques to find the core arguments in the document, followed by a page rank-based approach to select the most important arguments. Our domain of interest is similar in that different documents can present different viewpoints. However, in our case the information need of the user is to understand documents' underlying viewpoints, not the arguments that led to that viewpoint.
In addition, argument mining techniques are not suitable for viewpoint mining due to their different linguistic structure.

%% file: user_study.tex
\vspace{-2mm}
\section{Snippets' Reliability  in Commercial Search Engines}
\label{sec:study}
In the introduction section, we showed an example snippet from Google's search result, which, although relevant to the query phrase, could easily mislead a user. 
To gain a deeper understanding about the potential prevalence of such a phenomenon, we need to explore a wider variety of snippets and queries, as well as collect feedback from actual users. 

To that end, we designed a user study protocol in which participants were presented with a SERP discussing the effectiveness of a medical intervention in treating a medical condition.
The participants were requested to evaluate the viewpoint of the documents in the SERP, based only on their titles and snippets.
A SERP could contain snippets that were extracted either by a human annotator or from the Google search engine, which counts on a query-based  snippet generation method \cite{GoogleSnippets}.
We chose to use the Google search engine in our study due to its dominance in the industry \cite{davies2018meet}.
We then used participants' input to compare the reliability of snippets extracted by each method. 
In this context, a reliable snippet refers to one where, if a user is questioned about the document's viewpoint on the treatment's effectiveness, they would give the same response whether they had read the snippet or the complete document.
In order for this to occur, a reliable snippet needs to: 1) present a viewpoint about the effectiveness of the intervention mentioned in the query; and 2) provide the same viewpoint as that of its corresponding document. 
An unreliable snippet can potentially misinform a user, and therefore should be avoided in a SERP. 
\vspace{-4mm}
\subsection{User Study Design}
\label{sec:dataset}

\subsubsection{Dataset Preparation}
\label{sec:query_collection}
Before beginning our user study, we constructed a dataset of web documents and their associated snippets. The snippets were either provided by the Google search engine or extracted manually 
in order to provide a base for comparison in our experiment.

\textbf{Selecting Interventions and Conditions.}
All snippets in our study were extracted from web documents related to queries about the effectiveness of a medical intervention in treating a medical condition. We utilized the Cochrane reviews dataset to find interventions and conditions that would yield interesting search results.
A Cochrane review for a given medical intervention and condition is a process in which a team of professionals manually select and examine a set of random clinical trials discussing the effectiveness of the intervention in treating the condition, then publish a report with their conclusions 
\cite{cipriani2011cochrane,white2014content}. The Cochrane review dataset is available to the public~\cite{Cochrane}.
We randomly selected a set of reviews from this dataset in which the intervention was not overly complex for laymen readers to understand. 

After selecting a set of reviews, we further narrowed down the selection by choosing a random subset of reviews in which the true effectiveness of the intervention was either negative or inconclusive due to mixed results or insufficiently large-scale experiments. 
We made this selection based on the findings of a preliminary exploratory experiment, where we issued queries related to intervention-condition pairs (e.g., ``roselle for hypertension'') and analyzed the viewpoints in retrieved documents. This experiment revealed that when the intervention was effective in treating the condition, the retrieved documents expressed positive viewpoints almost exclusively, while ineffective or uncertain interventions elicited a wider range of viewpoints. Therefore, to ensure diversity in our dataset, we chose to focus on intervention-condition pairs with negative or inconclusive effectiveness.

\textbf{Query Phrasing and Document Retrieval.}
For each selected intervention-condition pair (e.g., roselle-hypertension), we then retrieved a set of relevant web documents and their associated snippets using Google. 

Search engine users can employ various forms of phrasing to inquire about the effectiveness of intervention X in treating a medical condition Y. Commonly used  phrasings include ``Is X helpful in treating Y?'' and ``Does X help Y?''. Negatively phrased queries such as ``Is X ineffective in treating Y?'' are also possible. However, research indicates that positive phrasing is the most frequently employed form when testing a hypothesis \cite{wason1960failure}.

In a query-based snippet extraction method, the way a query is phrased can impact the snippet extraction results. This is because all terms in the query, excluding stop words, are taken into consideration \cite{brin1998anatomy,GoogleSnippets}. Consequently, terms like ``effective'' and ``helps,'' as well as terms like ``ineffective'' or ``useless,'' can influence the selection of text snippets. In our initial study, we aimed to avoid introducing any positive bias that might stem from users' inclination to phrase their queries positively. Similarly, we refrained from introducing any negative bias by avoiding the use of negative phrases.

Therefore, during our document and snippet retrieval process, we formulated all queries in the neutral form of ``X for Y'' (for example, ``Roselle for hypertension''). This approach was employed to prevent any bias in the snippet extraction process. In Section \ref{sec:google_perf}, we will discuss the implications of this decision when presenting the study results for the query-based method.

\textbf{Viewpoint Labeling.}
After the documents were retrieved, they were manually labeled according to their viewpoints about the effectiveness of X in treating Y. 
A document's viewpoint could have one of four possible labels: 
\vspace{-1mm}
\begin{itemize}
    \item Ineffective -  X is not helpful in treating Y.
    \item Effective  -  X is helpful in treating Y.
    \item Inconclusive - A conclusion about the effectiveness of X in treating Y cannot be unilaterally determined.
    \item No Viewpoint - The document did not explicitly express any viewpoint as to the effectiveness of X in treating Y. 
\end{itemize}
\vspace{-1mm}
Each document was read and labeled by two paid annotators.  
The annotators were individuals with education in science and healthcare domains. Both had experience in reading content relevant to our study. Informed consent was obtained from both and approved by our institutional review board (IRB). The annotators did not have any contact with each other. 

Similarly to previous work \cite{white2014content,hashavit2021understanding}, to help the annotators with the nuances found in some of the documents, we provided them with two additional labeling options:
\textit{Potentially Effective} and  \textit{Potentially Ineffective}. The annotators were directed to use these labels when a document expressed some form of reservation as to its conclusion. For example, claiming that the intervention is effective only for specific cases, or that the conclusions are preliminary. 

Once all of the documents were labeled, the labels were regrouped to the previously defined four-level ratings (effective, ineffective, inconclusive, and no viewpoint) by unifying the \textit{potentially effective} label with the \textit{effective} label and the \textit{potentially ineffective} label with the \textit{ineffective} label.

\textbf{Manual Snippet Extraction.} One of the annotators was also responsible to extract snippets from documents, which would later be used as the ground-truth for comparison. 
This annotator's task was to identify a few sentences in the document that summarizes the viewpoint of the document regarding the effectiveness of X in treating Y.
The annotator was instructed to find a few sentences that could be used as a snippet. No hard length limit was provided to the annotator to avoid overcomplicating her task. If the snippet was too long, it was later cropped to match a uniform snippet length restriction for all examined methods.
An example of text snippets provided for a document discussing the treatment of asthma using acupuncture \cite{AcuAsthma} can be seen in Table \ref{tab:snippet}. The first two columns correspond to the methods compared in this section, and the last column corresponds to the algorithmic framework that will be discussed in Section  \ref{sec:alg}.

\begin{table*}[h]
\caption{An example of snippets extracted from the same document by the different extraction methods.}
\begin{tabular}{ |p{0.3\linewidth}|p{0.3\linewidth}| p{0.3\linewidth}|} 
 \hline
 \textbf{Google} & \textbf{Manual} & \textbf{Information Need-Focused Approach}\\
 \hline
  25 Jan 1999 - Acupuncture is a treatment originating from traditional Chinese medicine. It consists of the stimulation of defined points on the skin ...& There is not enough evidence to make recommendations about the value of acupuncture in asthma treatment. Further research needs to consider the complexities and...
 &there is insufficient evidence to make recommendations about the value of acupuncture as a treatment for asthma based on current evidence...\\
\hline
\end{tabular}
\label{tab:snippet}
\end{table*}

\textbf{Construction of the Final Dataset: Filtering and Query Selection.}
In the final dataset construction phase we filtered out the following documents:
\begin{itemize}
    \item Documents on which the two annotators did not agree on the four-level rating label.
    \item Documents that were labeled with the ``no viewpoint'' option. 
    \item Documents whose viewpoint regarding the intervention was declared in the title (since for these documents the snippets' contribution in users' decision making is negligible).
\end{itemize}

We then grouped all of the documents based on their respective intervention-condition pairs. In order to ensure that participants encountered diverse viewpoints during the study, each intervention-condition pair had to have a minimum of six documents associated with at least two different viewpoints in order to be included in the study. Intervention-condition pairs that did not meet these criteria were also excluded from the study.
As a result, our final study dataset comprised 42 documents, discussing 7 distinct intervention-condition pairs, with 6 documents dedicated to each query. Among these documents, 15 were labeled as effective in relation to their corresponding intervention-condition pair, 12 were labeled as inconclusive, and 15 were labeled as ineffective. This balanced distribution allowed for comprehensive exploration of different perspectives within the dataset.
The intervention and conditions included in the final dataset were as follow:

\begin{enumerate}
    \item \textit{Acupuncture for chronic asthma}.
    \item \textit{Acupuncture for epilepsy}.
    \item \textit{Aromatherapy for dementia}.
    \item \textit{Fermented milk for hypertension}.
    \item \textit{Ginkgo Biloba for tinnitus}.
    \item \textit{Glutamine for Crohn's disease}.
    \item \textit{Roselle for hypertension}.
\end{enumerate}

\subsubsection{SERP Generation}
\label{sec:serp_gen}
We define the combination of a SERP result's title and snippet as the result's caption. The participants' task was to assess the viewpoint of documents based on their captions. We therefore constructed a set of web pages that resembled a SERP in its conventional design in commercial search engines, e.g., Google. Each page contained a query box with the query text and a list of six result captions, corresponding to at least two different viewpoints.

Although the queries we used to retrieve the web articles were phrased neutrally, in the query box of SERPs presented to participants the phrasing was intentionally made positive (``Is roselle effective in treating hypertension?''). This design is because a positive framing is the popular phrasing form when testing a hypothesis using a search engine \cite{wason1960failure}, and we wanted the participants' experience to be as similar as possible to an actual web search.

The links to the actual documents were
disabled to prevent participants from accidentally visiting the websites.
Each SERP page corresponded to one query and all snippets in it were extracted using the same snippet extraction method (Google's or our manual extraction). Snippets  were limited to 160 characters regardless of their extraction method. 
In particular, the title in each caption was taken from Google's search result page and was kept identical for both snippet extraction methods. The order in which document captions were presented on a page was set randomly under each query to eliminate position bias. This order was identical for all documents on SERPs corresponding to the same query, regardless of the snippet extraction method. 


\subsection{User Study Procedure}
\label{sec:study_proc}
\begin{figure}[t]
\centering
\includegraphics[scale=0.45,keepaspectratio]{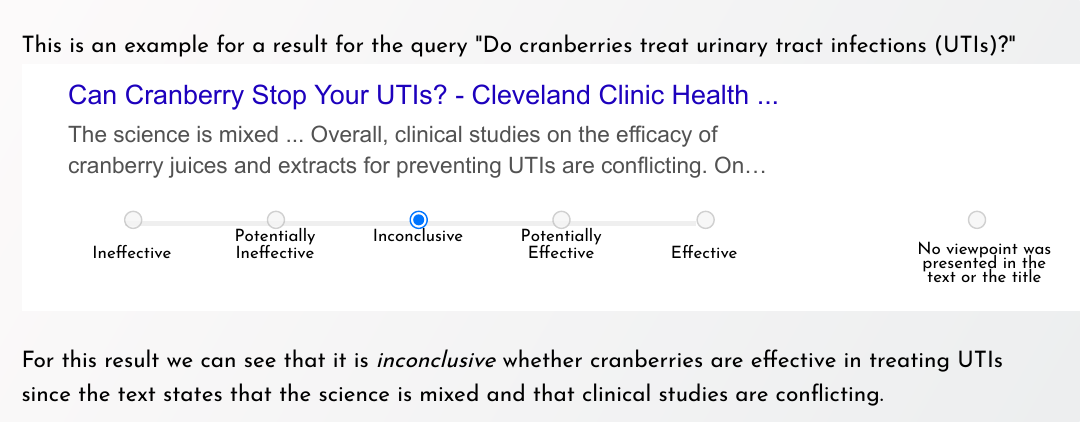}
\vspace{-3mm}
\caption{An example of the annotation instructions. }
\label{fig:instructions}
\vspace{-5mm}
\end{figure}

Our user study was conducted using the Amazon Mechanical Turk (MTurk) platform. We used the external survey option which directs participants to an external web page where the survey was implemented. Since the survey was in English, participants were restricted to English speaking countries only. 

\textbf{Explaining the task to participants.} Upon beginning the survey, each participant was requested to provide their demographics. 
Participants were then presented with a short explanation about the nature of the task.

The labeling options provided to participants were the same as those provided to our annotators, as described in Section \ref{sec:dataset}. 
To ensure that participants understood how to label the captions, after viewing the instructions, participants underwent a short tutorial. For each label option we provided an example of a result caption along with its correct label and a short explanation of why this was the correct option. An example given to participants in order to explain the \textit{Inconclusive} label option is provided in Figure \ref{fig:instructions}.

\textbf{Assuring the quality of workers}. In order to filter out unqualified workers, participants were also required to take a short test before continuing to the main survey.
Participants who completed the tutorial were directed to the test survey page.
In this test, they were requested to label the viewpoints expressed in a set of previously annotated captions. The test survey page was designed using the same setting as the actual annotation task.  


After the participants annotated all of the entries, their answers were regrouped to a four-level rating, as was done to the annotators' labels in Section \ref{sec:dataset}. The answers were then compared to the ground-truth labels. Participants who labeled all captions correctly were immediately directed to the main survey page, where
they were assigned a SERP with a chosen query and a designated snippet extraction method. None of the captions used in the tutorial or test overlap with our main study query dataset. 

\textbf{The annotation task.} The main survey page was designed 
according to the setting described in Section \ref{sec:serp_gen}.
The participants were then requested to label each caption. The participants could revisit the tutorial at any point by pressing a dedicated button.

\textbf{Post-survey question form.} Once all captions had been labeled, participants could submit that part of the survey and continue to the final question form. On this form, they were asked to evaluate the effectiveness of the intervention mentioned in the query, based on the captions they read on the previous page. 
They were given the same labeling options as in the caption labeling task, with one difference:  
Instead of the \textit{``No Viewpoint''} option provided in the caption labeling task, a \textit{``Not Sure''} option was given. Its phrasing was as follows: \textit{``Not sure, there was not enough information for me to reach a conclusion''}. This was done due to the difference in the meaning behind the two labels: In the snippet annotation task, participants were asked to annotate the viewpoint of a single snippet. The ``No Viewpoint'' phrase was therefore best suited. In the post-question form,  however, we asked the participants' opinion about the effectiveness of the intervention in general. For this question, the phrasing ``Not Sure'' was more suitable, since its intent is to describe a situation where the participant cannot form an opinion.

In addition to providing a label, participants were requested to note the level of confidence they felt in selecting their answer. The confidence levels ranged from 0 to 10, where 0 meant: \textit{``Extremely unconfident''} and 10 meant \textit{``Extremely confident''}. 
To get more insight, we additionally requested that participants write a few words about what made them select their answers.


\begin{table}[t]
\caption{Summary of the statistics of the user study dataset.}
\vspace{-1mm}
  \begin{tabular}{|l|c|}
   \hline
    \# of queries & 7 \\
    \hline
    \# of captions in a SERP & 6 \\
    \hline
    \# of responses per SERP & 10  \\
    \hline
    \# Total of snippets annotated per extraction method & 420 \\
  \hline
\end{tabular}
\label{tab:doc_num}
\vspace{-5mm}
\end{table}

\textbf{The completion page.} Participants who completed the survey or who failed the test were directed to the survey completion page to receive a verification code for claiming their payment from the MTurk website. On this page, they were also presented with a highlighted warning message noting that they were participating in an experiment and should not make any real conclusions about the true effectiveness of the intervention discussed in the study.
After this point, the participant could not retake the survey, not even for another survey in our study, thus ensuring that each survey corresponded to a single qualified participant.
Participants were paid \$0.2 for taking the test. If they passed the test they were paid an additional \$1.3 for their participation in the complete survey.


\textbf{Study validation.} For each snippet extraction method, 42 different snippets were annotated. Each snippet was annotated by 10 different participants. Excluding the ``No Viewpoint'' annotation option, annotations were measured on a 5-point Likert scale ranging from 1 (ineffective) to 5 (effective). 
To assess the average variability between annotations on a single snippet, we computed the standard deviation for each snippet's annotation set. 
This produced 42 values, each corresponding to a single snippet and its 10 annotations.
We then computed the mean values of each method's 42 standard deviation values. The mean standard deviation value was 0.592 for snippets extracted using the manual approach, and 0.665 for snippets extracted using the query-based approach.

This low variability indicates that participants tended to annotate individual snippets consistently when presented with the same extraction method. This, in turn, suggests that participants understood the instructions provided and performed their task diligently.

%% file: google_perf.tex

\subsection{User Study Results}
\label{sec:google_perf}

\begin{table}[t]
    \centering%
    \caption{Captions annotated as not presenting a viewpoint for the query-based and manual extraction methods. } \label{tab:NS}
        \begin{tabular}{|c|c|c|c|}
            \cline{2-4}
            \multicolumn{1}{c|}{}  & Query-based & Manual & \ \pbox{20cm}{ \# Annotated \\ Snippets} \\
            \hline
            \textbf{Effective} & 55 (36.6\%) & 18 (12\%) & 150\\
            \hline
            \textbf{Inconclusive} & 26 (21.6\%) & 15 (12.5\%) & 120  \\
            \hline 	
            \textbf{Ineffective} & 37 (24.6\%) & 5 (3.3\%) & 150\\
            \hline
            \hline
            \textbf{Total} & 118 (28\%) & 38 (9\%) & 420 \\
            \hline
        \end{tabular}
        \vspace{-4mm}
  \end{table}


\begin{table*}[t]
  \begin{subtable}[t]{.5\linewidth}%
    \centering%
    \caption{Query-Based } \label{tab:conf_Google}
        \begin{tabular}{|c|c|c|c|}
            \cline{2-4}
            \multicolumn{1}{c|}{}  &
            \pbox{20cm}{\textbf{Annotated} \\ \textbf{Effective}} & 
            \pbox{20cm}{\textbf{Annotated} \\ \textbf{Inconclusive}} & 
            \pbox{20cm}{\textbf{Annotated} \\ \textbf{Ineffective}} \\
            \hline
            \textbf{Effective} & 65 (68.4\%) & 22 (23.2\%) & 8 (8.4\%)\\
            \hline
            \textbf{Inconclusive} & 55 (58.5\%) & 29 (30.9\%) & 10 (10.6\%) \\
            \hline 	
            \textbf{Ineffective} &41(36.3\%) &  11(9.7\%)&  61(54\%) \\
            \hline
        \end{tabular}
  \end{subtable}%
  \begin{subtable}[t]{.5\linewidth}%
  \centering%
   \caption{Manual} \label{tab:conf_human}
        \begin{tabular}{|c|c|c|c|}
            \cline{2-4}
            \multicolumn{1}{c|}{}  &
            \pbox{20cm}{\textbf{Annotated} \\ \textbf{Effective}} & 
            \pbox{20cm}{\textbf{Annotated} \\ \textbf{Inconclusive}} & 
            \pbox{20cm}{\textbf{Annotated} \\ \textbf{Ineffective}} \\
            \hline
            \textbf{Effective} &122(92.4\%) & 5(3.8\%) & 5(3.8\%)\\
            \hline
            \textbf{Inconclusive} &4(3.8\%) & 71(67.6\%)   & 30(28.6\%) \\
            \hline
            \textbf{Ineffective} & 2 (1.4\%) & 6(4.1\%)   & 137(94.5\%)  \\
            \hline
        \end{tabular}
  \end{subtable}%
\caption{Participants' annotations of documents' viewpoints according to their captions.}
\label{tab:conf}
\vspace{-7mm}
\end{table*}

\textbf{Descriptive statistics about the study.} In total, 140 different MTurk workers participated in the complete survey.  The participants' ages ranged from 20 to 68, with a mean value of 40 (sd=11) and a fairly balanced gender distribution (53.3\% men, 46.2\% women). 
Table \ref{tab:doc_num} summarizes the dataset statistics for each snippet extraction method. 

\noindent\textbf{Failure to present a viewpoint}. Table \ref{tab:NS} reports, among all the captions labeled as ``No Viewpoint'' (i.e., 118+38), how many captions are related to documents labeled as effective, inconclusive and ineffective. The first two columns denote the extraction method and the last column details the total number of snippets annotated for each possible viewpoint. The number of annotated snippets per viewpoint is the same under both methods.
For example, the number of no viewpoint captions extracted from documents labeled as expressing an effective viewpoint is 55 for the query-based method and 18 for the manual one; these 55 snippets represent 36.6\% of total snippets related to documents labeled as effective, i.e.,~150.

Inspecting the table we observe that for the query-based extraction method 118 (28\%) out of the 420 snippets were annotated by participants as not presenting a viewpoint. Meaning that in over a quarter of the cases participants could not infer the viewpoint presented in the document from its snippet, as opposed to only 38 (9\%) out of the 420 snippets extracted by a human reader. 
By inspecting the different viewpoints, we see that this issue is not unique for a specific viewpoint. In fact, in all of the viewpoints we observe a higher percentage of documents labeled as ``No Viewpoint'' for snippets extracted by the query-based method compared to the manual method. 

\noindent\textbf{Snippets' annotation in relation to their corresponding documents}. We now continue to the second condition of snippet accuracy, i.e., whether it expresses the same viewpoint as its corresponding document. 
Tables \ref{tab:conf_Google} and \ref{tab:conf_human} present participants' annotations, excluding the ``No Viewpoint'' annotation option, for the query-based and the manual extraction methods, respectively. The rows correspond to viewpoint labels of documents, as labeled by our external annotators, and the columns correspond to the documents' snippet annotations given by participants. 
The values outside of the parentheses in each cell indicate the number of captions annotated with the corresponding column label for documents classified according to the respective row label.
The values inside the parentheses denote the portion of these numbers in relation to all documents with the same viewpoint label. The values in the parentheses in each row sum up to 100.

We refer to snippets annotated with the same label as their corresponding documents as accurately annotated snippets.
When the snippets were extracted manually, 330 snippets were annotated accurately, representing 79\% of all annotated snippets and 86\% of all snippets annotated as expressing a viewpoint (i.e., excluding the ``No Viewpoint'' annotations) for this method. 
For the query-based method, we observe much lower results with only 155 snippets annotated according to their corresponding documents', representing 37\% of all annotated snippets and 51\% of all snippets annotated as expressing a viewpoint. This suggests that in the query-based method almost half of the snippets, annotated with a viewpoint, were annotated inaccurately.
To establish the significance of these observations, we conducted two $\chi^2$ tests to compare the proportions of accurately annotated snippets between the two methods. These comparisons were made with respect to snippets annotated as having a viewpoint, 
 as well as with respect to all snippets. 
Notably, all the comparisons yielded highly significant results ($p < 0.0001$).


Comparing the accuracy of both methods for each viewpoint label separately, we observe a significant difference in the portion of accurate annotations, as well. 
The manual extraction method outperforms the query-based approach by 24\% for documents whose viewpoint was labeled as effective, 36.7\% for documents labeled with an inconclusive viewpoint label, and 40.5\% for documents with an ineffective viewpoint label.
Once again, in order to validate the significance of these observations, a series of $\chi^2$ tests of independence were conducted. These tests compared the annotation proportions between the two snippet extraction methods 
for each document viewpoint 
(e.g. the first row of Table \ref{tab:conf_Google} and Table \ref{tab:conf_human}). All of the tests produced a significant p-value ($p < 0.001$).
This validates that the 
proportion of snippet annotations differed
significantly between the 
two methods, for all document viewpoints.


Inspecting Table \ref{tab:conf_Google} more closely,  we additionally observe a positive bias in the snippets extracted by the query-based method. This phenomenon is most severe when the related document's viewpoint is inconclusive, where participants are inclined to misinterpret the document's viewpoint as supporting the intervention's effectiveness in 58.5\% of cases. However, it is also present for documents labeled with the ineffective label, where over a third of snippets were incorrectly annotated as having an effective viewpoint. 

As mentioned in Section \ref{sec:dataset}, the snippet retrieval process for the query-based method was done using a neutrally phrased query. The most popular phrasing form when testing a hypothesis using a search engine, however, is positive phrasing such as ``Is X effective in treating Y'' or ``Does X help Y'' \cite{wason1960failure,white2013beliefs}. We believe this phrasing form will further exacerbate the positive bias in the query-based method, due to the existence of positive terms in the query phrase.
We leave the validation of this belief as interesting future work.

\noindent\textbf{Participants' perceived answers.}  After participants annotated the captions, they were questioned about their belief regarding the effectiveness of the intervention for the condition discussed in the query, based on the captions they read. We refer to this answer as the participant's perceived answer. The vast majority of participants were able to provide an answer, with only 6 out of 70 participants (8.5\%) choosing the ``Not sure'' option for the query-based method, versus 0 participants for the manual extraction method. 
Participants were also required to provide the confidence level regarding their answer.
The confidence levels of participants in their answers were similar for both extraction methods, with a mean confidence level of 6.8 for the manual method and a value of 7 for the query-based method. These confidence levels are high despite the fact that participants read only captions and could not visit the actual websites. Upon inspecting the reasons that they provided, we noticed that participants frequently referred to the information found in the snippets as a basis for their decisions.
These findings indicate that users can form confident opinions on health-related topics based on the information displayed in search engine results pages, even without visiting the actual sources. This behaviour makes them susceptible to snippets that misrepresent their corresponding documents.

\noindent\textbf{The origin of observed behaviours and a possible solution.}
We believe that the origin of the issues observed in this section stems from the fact that generating accurate snippets for this specific information need requires an understanding of both the user's intent and that of the collected documents. However, the current query-based snippet extraction process relies on matching query terms with document text, without comprehending the core intentions of the users or the retrieved documents. Consequently, in our scenario, keyword-focused snippets may misrepresent the corresponding documents, which can then mislead the users.

Naturally, we cannot expect an automatic method to outperform humans in such a complex task. However, given the encouraging progress in related text extraction tasks and the significant performance gaps observed in this study, we believe that solutions that are more attuned to the user's core information need in these types of queries will produce better results. 

We are aware that in order for this approach to be usable, the search engine must be able to identify the information need from the query phrase. 
This task is still non-trivial in a large scale. It is, however, possible to implement under specific contexts, and we do see it in commonly provided features, such as direct answers, 
where the search engine directly provides users with answers regarding their searches for weather reports or stock information.
It is also possible to implement such a specialized procedure in a health-related search for queries such as intervention effectiveness queries, whose structure is relatively easy to detect (i.e., X for Y or X treats Y). 
Once the user's information need is well-defined, a dedicated snippet extraction method can be employed to extract more reliable snippets.

Given that the stake of misinformation in this domain can never be underestimated, we believe that the need for developing such a procedure is self-evident.
In the next section, we will present one solution implementation for an information need-focused snippet extraction method for intervention effectiveness queries.

%% file: algorithm.tex
\vspace{-2mm}
\section{Information Need-Focused Snippet Extraction}
\label{sec:alg}
In this section, we present an information need-focused snippet extraction solution for queries discussing medical interventions' effectiveness. Section \ref{sec:method} will detail the technical aspects of this solution and Section \ref{sec:alg_perf} will compare its performance to that of the query-based and  manual snippet extraction approaches. 

\subsection{Classification Driven Snippet Extraction}
\label{sec:method}
\textbf{Motivation.} 
We consider a relevant document for answering an intervention effectiveness query to be one that provides a clear viewpoint on the effectiveness of the specific intervention and condition mentioned in the query.
Correspondingly, a reliable snippet for such a document is a fragment of that document's text which summarizes the viewpoint of the document regarding the intervention's effectiveness. Notice that in one document there could be several sentences expressing a viewpoint regarding the intervention's effectiveness, which are not \textit{the document's} viewpoint regarding the intervention's effectiveness (recall the example discussed in our introduction, Figure \ref{fig:example}). An automatic snippet extraction solution must be able to distinguish between the two.

Our proposed solution consists of two components combined into a single framework. When given a document, intervention, and condition, the framework initially infers the document's viewpoint about the effectiveness of the intervention for the given condition. It then selects the text fragment that, if removed, will have the most significant impact on the classification result. This technique is sometimes referred to as representation erasure. It was originally proposed as a means to explain neural networks' decisions \cite{li2016understanding}.

\noindent\textbf{Formal Definition.}
Let $s_d^1, s_d^2,..,s_d^n$ be the sequence of sentences whose concatenation results in a document $d$. Let $M$ be a trained prediction model that, given a document $d$, an intervention-condition pair $IC$, and a set of possible viewpoints $V$, outputs a score $M_v(d,IC)$ for each possible viewpoint such that $v^* = \operatorname*{arg\,max}_v M_v(d, IC)$ is the viewpoint 
classification for document $d$ concerning intervention-condition pair IC.
We denote by $d_{-i}$ the sequence of sentences composing document $d$ with sentence $s_d^i$ omitted from the sequence. 
The contribution of sentence $s_d^i$ to $M_v(d,IC)$ is defined as: 
\begin{equation} \label{eq:cont}
C_M(s_d^i, v, IC) = M_v(d, IC) - M_v(d_{-i}, IC) 
\end{equation}

We then extract the sentence with the highest contribution to the document's viewpoint 
classification as our selected snippet: $\operatorname*{arg\,max}_s  C_M(s, v^*, IC) $.

\subsubsection{Framework Implementation}
\hfill\\
\textbf{First Component: Viewpoint Classification Model}.
To develop a viewpoint classification model, we created a BERT-based classifier. BERT (Bidirectional Encoder Representations from Transformers) is a machine learning technique for natural language processing that has been widely used for various NLP tasks \cite{devlin2018bert}. The BERT encoder outputs an embedding of the input text, which can be fine-tuned for specific NLP tasks through a relatively simple fine-tuning phase. In our study, since we focused on the medical domain, we utilized BioBERT \cite{lee2020biobert}, a version of BERT that is specifically trained on medically related data. 

To fine-tune the model for our viewpoint classification task, we trained it to predict the viewpoints of documents. The dataset used to train the classifier was obtained from a previous study conducted by Hashavit et al. \cite{hashavit2021understanding}, which contained 2,837 medical paper abstracts discussing 287 intervention-condition pairs. Since the vast majority of documents in this dataset discussed a single intervention-condition pair, we did not need to include the intervention and condition as input to the classifier. This simplified the classification process, resembling a multi-class classification task.
We used the BioBERT encoder followed by one linear layer to transfer the encoder output to our required three-class output (i.e., effective, ineffective, and inconclusive). Further details on the fine-tuning process for BioBERT can be found in \cite{lee2020biobert}.

\noindent\textbf{Second Component: Snippet Extraction}.
After training the viewpoint classification model, we used it to extract snippets from documents. However, since the classifier was trained on short documents (i.e., paper abstracts) and did not receive the intervention-condition pair as input, we needed to run a pre-processing phase to filter out irrelevant data and noise from the document before feeding it into the classifier.

To filter out noisy data, we created a sub-document that contained at most the following 5 paragraphs:
\begin{itemize}
    \item The first paragraph. 
    \item The last 3 paragraphs.
    \item The paragraph most relevant to the intervention and condition terms, using the Okapi BM25 metric \cite{robertson1995okapi}.
\end{itemize}

These paragraphs were chosen heuristically in an attempt to locate text segments that are most likely to contain the document's viewpoint summaries, following an intuition that summaries are usually presented either at the beginning or towards the end of documents. We additionally included the paragraph most relevant to the intervention and condition terms, in order to cover cases where the summary did not appear in any of these paragraphs. 
We refer to the output of the paragraph filtering phase as $d'$.

Next, we employed  another preprocessing step meant to remove the portions of $d'$ irrelevant to its corresponding intervention-condition pair.
In this phase $d'$ was partitioned to windows of size 510 words. A window that did not contain both the intervention and condition terms was filtered out. This filtering phase resulted in another sub-document $d''$. 

After concluding the preprocessing phase, we proceeded to acquire the viewpoint classification of $d''$ by utilizing our viewpoint classifier. We computed the contribution of each sentence to $d''$'s viewpoint classification, and selected the sentence with the highest contribution score.
An example of a snippet provided by the framework can be seen in the last column of Table \ref{tab:snippet}.

\begin{table}[b]
\vspace{-4mm}
    \centering%
    \caption{Captions extracted by our framework and annotated as not presenting a viewpoint.} 
    \label{tab:NS_Alg}
        \begin{tabular}{|c|c|c|}
            \cline{2-3}
            \multicolumn{1}{c|}{}  & \pbox{20cm}{ Information Need \\ Focused Approach} & \pbox{20cm}{ \# Annotated \\ Snippets}  \\
            \hline
            \textbf{Effective} & 19(12.67\%) & 150\\
            \hline
            \textbf{Inconclusive} & 31(25.83\%) & 120 \\
            \hline 	
            \textbf{Ineffective} & 33(22\%) & 150\\
            \hline
            \hline
            \textbf{Total} & 83(19.76\%)  & 420\\
            \hline
        \end{tabular}
        \vspace{-5mm}
  \end{table}

\subsection{Evaluation}
\label{sec:alg_perf}
We utilized our framework to extract snippets for the 42 documents that were extracted in Section \ref{sec:dataset} and subsequently conducted a user study to evaluate the reliability of these snippets.
This study was also conducted using the Amazon Mechanical Turk platform, following the same design and procedure as described in Section \ref{sec:dataset} and \ref{sec:study_proc}, with snippets extracted using our framework. We used the same statistical analysis tools to evaluate the framework's performance as in Section \ref{sec:google_perf}.
Of course, participants who
participated in the user study for the other two methods were not eligible to
participate in this study.

The demographics of participants in this study were very similar to those of the previous study for all categories  except in the gender distribution, which showed a slight imbalance in comparison to the two other methods (45\% men, 55\% women). We therefore compared the distribution of participants' responses between men and women, and their responses showed no statistical difference under any of the snippet extraction methods. 
What's more, the mean standard deviation reported in Section \ref{sec:study_proc} for study validation was 0.52, a low value very similar to the values reported for the other two methods.

Table \ref{tab:NS_Alg} details the number of captions extracted by our method and annotated as not having a viewpoint, following the same manner as in Table \ref{tab:NS}. In total, out of the 420 annotated captions, 83 (19.76\%) captions were annotated as not having a viewpoint. While this is higher than the 9\% presented by manually extracted snippets, it is also significantly lower compared to the 28\% presented by the query-based method. Upon inspecting the different viewpoint labels, we additionally observe that the advantage of the framework is primarily evident when documents are assigned an effective viewpoint label. In such cases, a mere 12.67\% of the documents were labeled as lacking a viewpoint , which is quite similar to the 12\% exhibited by the manual method. In contrast, the query-based method exhibited a significantly higher percentage of 36.6\%. This significant discrepancy amounts to a difference of nearly 24\%.


Table \ref{tab:conf_alg} breaks down participants' annotations according to documents' viewpoints, similar to Tables \ref{tab:conf_Google} and \ref{tab:conf_human}, which did so for the other two methods.
In total, out of the 337 captions annotated with a viewpoint, 252 were annotated with the same viewpoint as their corresponding documents. This represents 74.7\% of captions annotated as having a viewpoint and 60\% of all snippet annotations for those extracted by our method. These values are both significantly higher than the 51\% and 37\% achieved by the query-based method 
and also significantly lower than the 86\% and 79\% achieved by the manually extracted captions (p < 0.00001). 
 Unlike the query-based method, the majority of captions extracted by our framework were accurately annotated by participants. 

Inspecting each viewpoint individually, we observe that the information need-focused method outperformed the query-based method for all documents' viewpoints.
This is especially evident when the document's viewpoint is ineffective. In this case, 82\% of captions were annotated according to their corresponding documents' viewpoints, compared to only 54\% for the query-based approach, which is an advantage of 28\% in favor of the information need-focused method.
The information need-focused method also outperforms the query-based method by 15.6\% for documents annotated with an effective viewpoint. Documents 
 with an inconclusive viewpoint were again the most challenging to extract an accurate snippet from. Such documents received the lowest accuracy of 52\%, but it is still 21.1\% higher than that of the query-based method, for which only 30.9\% of captions were annotated correctly.
As expected, manually extracted snippets outperformed the information need-focused method by 8.4\% and 12.5\% for documents labeled as having an effective or ineffective viewpoint, and by 15.6\% for documents labeled as having an inconclusive viewpoint. 
A series of $\chi^2$ tests of independence confirmed that the proportion of snippet annotations differed significantly between the information need-focused approach and each alternative method for all viewpoints (p < 0.01).

\begin{table}[t]
    \centering%
    \caption{Participants’ annotation of documents’ viewpoints according to captions extracted by our framework} \label{tab:conf_alg}
        \begin{tabular}{|c|c|c|c|}
            \cline{2-4}
            \multicolumn{1}{c|}{}  &
            \pbox{20cm}{\textbf{Annotated} \\ \textbf{Effective}} & 
            \pbox{20cm}{\textbf{Annotated} \\ \textbf{Inconclusive}} & 
            \pbox{20cm}{\textbf{Annotated} \\ \textbf{Ineffective}} \\
            \hline
            \textbf{Effective} & 110 (84\%)  & 19 (14.5\%)  & 2(1.5\%) \\
            \hline
            \textbf{Inconclusive} & 19(21\%)  & 46(52\%) & 24 (27\%) \\
            \hline 	
            \textbf{Ineffective} & 7 (6\%)& 14 (12\%) & 96(82\%)  \\
            \hline
        \end{tabular}
        \vspace{-7mm}
  \end{table}%



By inspecting participants' perceived answers, we observe that the results are similar to the query-based approach with only 4 participants choosing the ``Not Sure'' label. The confidence levels were also similar to the other two methods at 6.92. 

To summarize, our observations indicate that the information need-focused approach surpasses  the query-based approach both in 
presenting a viewpoint within the snippets and accurately reflecting the viewpoint of the snippets' corresponding documents. 
Consequently, we conclude that the information need-focused approach generates snippets that are significantly more reliable compared to the query-based approach.
This proves that an improvement in the snippet extraction process for viewpoint-related queries, such as intervention effectiveness queries, is definitely feasible. The information need-focused approach is still outperformed by the manual extraction method. However, the proposed framework is quite rudimentary and technically simple. A more sophisticated approach can surely further minimize this gap. 


%% file: main.bbl

\begin{thebibliography}{41}


\ifx \showCODEN    \undefined \def \showCODEN     #1{\unskip}     \fi
\ifx \showDOI      \undefined \def \showDOI       #1{#1}\fi
\ifx \showISBNx    \undefined \def \showISBNx     #1{\unskip}     \fi
\ifx \showISBNxiii \undefined \def \showISBNxiii  #1{\unskip}     \fi
\ifx \showISSN     \undefined \def \showISSN      #1{\unskip}     \fi
\ifx \showLCCN     \undefined \def \showLCCN      #1{\unskip}     \fi
\ifx \shownote     \undefined \def \shownote      #1{#1}          \fi
\ifx \showarticletitle \undefined \def \showarticletitle #1{#1}   \fi
\ifx \showURL      \undefined \def \showURL       {\relax}        \fi
\providecommand\bibfield[2]{#2}
\providecommand\bibinfo[2]{#2}
\providecommand\natexlab[1]{#1}
\providecommand\showeprint[2][]{arXiv:#2}

\bibitem[Acu(1999)]%
        {AcuAsthma}
 \bibinfo{year}{1999}\natexlab{}.
\newblock \bibinfo{title}{{Acupuncture for chronic asthma}}.
\newblock \bibinfo{howpublished}{\url{https://www.cochrane.org/CD000008/AIRWAYS_acupuncture-for-chronic-asthma}}.
\newblock


\bibitem[Goo(2022)]%
        {GoogleSnippets}
 \bibinfo{year}{2022}\natexlab{}.
\newblock \bibinfo{title}{{Control your snippets in search results}}.
\newblock \bibinfo{howpublished}{\url{https://developers.google.com/search/docs/appearance/snippet}}.
\newblock


\bibitem[Coc(2023)]%
        {Cochrane}
 \bibinfo{year}{2023}\natexlab{}.
\newblock \bibinfo{title}{{Cochrane Library: Cochrane Reviews}}.
\newblock \bibinfo{howpublished}{\url{https://www.cochranelibrary.com/}}.
\newblock
\urldef\tempurl%
\url{https://www.cochranelibrary.com/}
\showURL{%
\tempurl}


\bibitem[Fea(2023)]%
        {FeaturedSnippets}
 \bibinfo{year}{2023}\natexlab{}.
\newblock \bibinfo{title}{{Featured snippets and your website}}.
\newblock \bibinfo{howpublished}{\url{https://developers.google.com/search/docs/appearance/featured-snippets}}.
\newblock


\bibitem[Ageev et~al\mbox{.}(2013)]%
        {ageev2013improving}
\bibfield{author}{\bibinfo{person}{Mikhail Ageev}, \bibinfo{person}{Dmitry Lagun}, {and} \bibinfo{person}{Eugene Agichtein}.} \bibinfo{year}{2013}\natexlab{}.
\newblock \showarticletitle{Improving search result summaries by using searcher behavior data}. In \bibinfo{booktitle}{\emph{Proceedings of the 36th international acm sigir conference on research and development in information retrieval}}. \bibinfo{pages}{13--22}.
\newblock


\bibitem[Alshomary et~al\mbox{.}(2020)]%
        {alshomary2020extractive}
\bibfield{author}{\bibinfo{person}{Milad Alshomary}, \bibinfo{person}{Nick D{\"u}sterhus}, {and} \bibinfo{person}{Henning Wachsmuth}.} \bibinfo{year}{2020}\natexlab{}.
\newblock \showarticletitle{Extractive snippet generation for arguments}. In \bibinfo{booktitle}{\emph{Proceedings of the 43rd International ACM SIGIR Conference on Research and Development in Information Retrieval}}. \bibinfo{pages}{1969--1972}.
\newblock


\bibitem[Bink et~al\mbox{.}(2023)]%
        {bink2023investigating}
\bibfield{author}{\bibinfo{person}{Markus Bink}, \bibinfo{person}{Sebastian Schwarz}, \bibinfo{person}{Tim Draws}, {and} \bibinfo{person}{David Elsweiler}.} \bibinfo{year}{2023}\natexlab{}.
\newblock \showarticletitle{Investigating the Influence of Featured Snippets on User Attitudes}. In \bibinfo{booktitle}{\emph{Proceedings of the 2023 Conference on Human Information Interaction and Retrieval}}. \bibinfo{pages}{211--220}.
\newblock


\bibitem[Bink et~al\mbox{.}(2022)]%
        {bink2022featured}
\bibfield{author}{\bibinfo{person}{Markus Bink}, \bibinfo{person}{Steven Zimmerman}, {and} \bibinfo{person}{David Elsweiler}.} \bibinfo{year}{2022}\natexlab{}.
\newblock \showarticletitle{Featured Snippets and their Influence on Users’ Credibility Judgements}. In \bibinfo{booktitle}{\emph{ACM SIGIR Conference on Human Information Interaction and Retrieval}}. \bibinfo{pages}{113--122}.
\newblock


\bibitem[Brin and Page(1998)]%
        {brin1998anatomy}
\bibfield{author}{\bibinfo{person}{Sergey Brin} {and} \bibinfo{person}{Lawrence Page}.} \bibinfo{year}{1998}\natexlab{}.
\newblock \showarticletitle{The anatomy of a large-scale hypertextual web search engine}.
\newblock \bibinfo{journal}{\emph{Computer networks and ISDN systems}} \bibinfo{volume}{30}, \bibinfo{number}{1-7} (\bibinfo{year}{1998}), \bibinfo{pages}{107--117}.
\newblock


\bibitem[Cipriani et~al\mbox{.}(2011)]%
        {cipriani2011cochrane}
\bibfield{author}{\bibinfo{person}{A Cipriani}, \bibinfo{person}{TA Furukawa}, {and} \bibinfo{person}{C Barbui}.} \bibinfo{year}{2011}\natexlab{}.
\newblock \showarticletitle{What is a Cochrane review?}
\newblock \bibinfo{journal}{\emph{Epidemiology and psychiatric sciences}} \bibinfo{volume}{20}, \bibinfo{number}{3} (\bibinfo{year}{2011}), \bibinfo{pages}{231--233}.
\newblock


\bibitem[Clarke et~al\mbox{.}(2007)]%
        {clarke2007influence}
\bibfield{author}{\bibinfo{person}{Charles~LA Clarke}, \bibinfo{person}{Eugene Agichtein}, \bibinfo{person}{Susan Dumais}, {and} \bibinfo{person}{Ryen~W White}.} \bibinfo{year}{2007}\natexlab{}.
\newblock \showarticletitle{The influence of caption features on clickthrough patterns in web search}. In \bibinfo{booktitle}{\emph{Proceedings of the 30th annual international ACM SIGIR conference on Research and development in information retrieval}}. \bibinfo{pages}{135--142}.
\newblock


\bibitem[Cutrell and Guan(2007)]%
        {cutrell2007you}
\bibfield{author}{\bibinfo{person}{Edward Cutrell} {and} \bibinfo{person}{Zhiwei Guan}.} \bibinfo{year}{2007}\natexlab{}.
\newblock \showarticletitle{What are you looking for? An eye-tracking study of information usage in web search}. In \bibinfo{booktitle}{\emph{Proceedings of the SIGCHI conference on Human factors in computing systems}}. \bibinfo{pages}{407--416}.
\newblock


\bibitem[Davies(2018)]%
        {davies2018meet}
\bibfield{author}{\bibinfo{person}{Dave Davies}.} \bibinfo{year}{2018}\natexlab{}.
\newblock \showarticletitle{Meet the 7 most popular search engines in the world}.
\newblock \bibinfo{journal}{\emph{Search Engine Journal}} (\bibinfo{year}{2018}).
\newblock


\bibitem[Devlin et~al\mbox{.}(2018)]%
        {devlin2018bert}
\bibfield{author}{\bibinfo{person}{Jacob Devlin}, \bibinfo{person}{Ming-Wei Chang}, \bibinfo{person}{Kenton Lee}, {and} \bibinfo{person}{Kristina Toutanova}.} \bibinfo{year}{2018}\natexlab{}.
\newblock \showarticletitle{Bert: Pre-training of deep bidirectional transformers for language understanding}.
\newblock \bibinfo{journal}{\emph{arXiv preprint arXiv:1810.04805}} (\bibinfo{year}{2018}).
\newblock


\bibitem[Diriye et~al\mbox{.}(2012)]%
        {diriye2012leaving}
\bibfield{author}{\bibinfo{person}{Abdigani Diriye}, \bibinfo{person}{Ryen White}, \bibinfo{person}{Georg Buscher}, {and} \bibinfo{person}{Susan Dumais}.} \bibinfo{year}{2012}\natexlab{}.
\newblock \showarticletitle{Leaving so soon? Understanding and predicting web search abandonment rationales}. In \bibinfo{booktitle}{\emph{Proceedings of the 21st ACM international conference on Information and knowledge management}}. \bibinfo{pages}{1025--1034}.
\newblock


\bibitem[Ghenai et~al\mbox{.}(2020)]%
        {ghenai2020think}
\bibfield{author}{\bibinfo{person}{Amira Ghenai}, \bibinfo{person}{Mark~D Smucker}, {and} \bibinfo{person}{Charles~LA Clarke}.} \bibinfo{year}{2020}\natexlab{}.
\newblock \showarticletitle{A think-aloud study to understand factors affecting online health search}. In \bibinfo{booktitle}{\emph{Proceedings of the 2020 conference on human information interaction and retrieval}}. \bibinfo{pages}{273--282}.
\newblock


\bibitem[Goel et~al\mbox{.}(2009)]%
        {goel2009introducing}
\bibfield{author}{\bibinfo{person}{Kavi Goel}, \bibinfo{person}{Ramanathan~V Guha}, {and} \bibinfo{person}{Othar Hansson}.} \bibinfo{year}{2009}\natexlab{}.
\newblock \showarticletitle{Introducing rich snippets}.
\newblock \bibinfo{journal}{\emph{Google Webmaster Central Blog}} (\bibinfo{year}{2009}).
\newblock


\bibitem[Hashavit et~al\mbox{.}(2021)]%
        {hashavit2021understanding}
\bibfield{author}{\bibinfo{person}{Anat Hashavit}, \bibinfo{person}{Hongning Wang}, \bibinfo{person}{Raz Lin}, \bibinfo{person}{Tamar Stern}, {and} \bibinfo{person}{Sarit Kraus}.} \bibinfo{year}{2021}\natexlab{}.
\newblock \showarticletitle{Understanding and Mitigating Bias in Online Health Search}.
\newblock  (\bibinfo{year}{2021}), \bibinfo{pages}{265–274}.
\newblock
\showISBNx{9781450380379}
\urldef\tempurl%
\url{https://doi.org/10.1145/3404835.3462930}
\showURL{%
\tempurl}


\bibitem[Hashavit et~al\mbox{.}(2022)]%
        {hashavit2022not}
\bibfield{author}{\bibinfo{person}{Anat Hashavit}, \bibinfo{person}{Hongning Wang}, \bibinfo{person}{Tamar Stern}, {and} \bibinfo{person}{Sarit Kraus}.} \bibinfo{year}{2022}\natexlab{}.
\newblock \showarticletitle{Not Just Skipping. Understanding the Effect of Sponsored Content on Users' Decision-Making in Online Health Search}.
\newblock \bibinfo{journal}{\emph{arXiv preprint arXiv:2207.04445}} (\bibinfo{year}{2022}).
\newblock


\bibitem[Hop et~al\mbox{.}(2010)]%
        {hop2010automatic}
\bibfield{author}{\bibinfo{person}{Walter Hop}, \bibinfo{person}{Stephan Lachner}, \bibinfo{person}{Flavius Frasincar}, {and} \bibinfo{person}{Roberto De~Virgilio}.} \bibinfo{year}{2010}\natexlab{}.
\newblock \showarticletitle{Automatic Web Page Annotation with Google Rich Snippets.}. In \bibinfo{booktitle}{\emph{OTM Conferences (2)}}. \bibinfo{pages}{957--974}.
\newblock


\bibitem[Kanungo and Orr(2009)]%
        {kanungo2009predicting}
\bibfield{author}{\bibinfo{person}{Tapas Kanungo} {and} \bibinfo{person}{David Orr}.} \bibinfo{year}{2009}\natexlab{}.
\newblock \showarticletitle{Predicting the readability of short web summaries}. In \bibinfo{booktitle}{\emph{Proceedings of the Second ACM International Conference on Web Search and Data Mining}}. \bibinfo{pages}{202--211}.
\newblock


\bibitem[Leal~Bando et~al\mbox{.}(2015)]%
        {leal2015query}
\bibfield{author}{\bibinfo{person}{Lorena Leal~Bando}, \bibinfo{person}{Falk Scholer}, {and} \bibinfo{person}{Andrew Turpin}.} \bibinfo{year}{2015}\natexlab{}.
\newblock \showarticletitle{Query-biased summary generation assisted by query expansion}.
\newblock \bibinfo{journal}{\emph{Journal of the Association for Information Science and Technology}} \bibinfo{volume}{66}, \bibinfo{number}{5} (\bibinfo{year}{2015}), \bibinfo{pages}{961--979}.
\newblock


\bibitem[Lee et~al\mbox{.}(2020)]%
        {lee2020biobert}
\bibfield{author}{\bibinfo{person}{Jinhyuk Lee}, \bibinfo{person}{Wonjin Yoon}, \bibinfo{person}{Sungdong Kim}, \bibinfo{person}{Donghyeon Kim}, \bibinfo{person}{Sunkyu Kim}, \bibinfo{person}{Chan~Ho So}, {and} \bibinfo{person}{Jaewoo Kang}.} \bibinfo{year}{2020}\natexlab{}.
\newblock \showarticletitle{BioBERT: a pre-trained biomedical language representation model for biomedical text mining}.
\newblock \bibinfo{journal}{\emph{Bioinformatics}} \bibinfo{volume}{36}, \bibinfo{number}{4} (\bibinfo{year}{2020}), \bibinfo{pages}{1234--1240}.
\newblock


\bibitem[Li et~al\mbox{.}(2009)]%
        {li2009good}
\bibfield{author}{\bibinfo{person}{Jane Li}, \bibinfo{person}{Scott Huffman}, {and} \bibinfo{person}{Akihito Tokuda}.} \bibinfo{year}{2009}\natexlab{}.
\newblock \showarticletitle{Good abandonment in mobile and PC internet search}. In \bibinfo{booktitle}{\emph{Proceedings of the 32nd international ACM SIGIR conference on Research and development in information retrieval}}. \bibinfo{pages}{43--50}.
\newblock


\bibitem[Li et~al\mbox{.}(2016)]%
        {li2016understanding}
\bibfield{author}{\bibinfo{person}{Jiwei Li}, \bibinfo{person}{Will Monroe}, {and} \bibinfo{person}{Dan Jurafsky}.} \bibinfo{year}{2016}\natexlab{}.
\newblock \showarticletitle{Understanding neural networks through representation erasure}.
\newblock \bibinfo{journal}{\emph{arXiv preprint arXiv:1612.08220}} (\bibinfo{year}{2016}).
\newblock


\bibitem[Li and Chen(2010)]%
        {li2010personalized}
\bibfield{author}{\bibinfo{person}{Qing Li} {and} \bibinfo{person}{Yuanzhu~Peter Chen}.} \bibinfo{year}{2010}\natexlab{}.
\newblock \showarticletitle{Personalized text snippet extraction using statistical language models}.
\newblock \bibinfo{journal}{\emph{Pattern Recognition}} \bibinfo{volume}{43}, \bibinfo{number}{1} (\bibinfo{year}{2010}), \bibinfo{pages}{378--386}.
\newblock


\bibitem[Lippi and Torroni(2016)]%
        {lippi2016argumentation}
\bibfield{author}{\bibinfo{person}{Marco Lippi} {and} \bibinfo{person}{Paolo Torroni}.} \bibinfo{year}{2016}\natexlab{}.
\newblock \showarticletitle{Argumentation mining: State of the art and emerging trends}.
\newblock \bibinfo{journal}{\emph{ACM Transactions on Internet Technology (TOIT)}} \bibinfo{volume}{16}, \bibinfo{number}{2} (\bibinfo{year}{2016}), \bibinfo{pages}{1--25}.
\newblock


\bibitem[Maxwell et~al\mbox{.}(2017)]%
        {maxwell2017study}
\bibfield{author}{\bibinfo{person}{David Maxwell}, \bibinfo{person}{Leif Azzopardi}, {and} \bibinfo{person}{Yashar Moshfeghi}.} \bibinfo{year}{2017}\natexlab{}.
\newblock \showarticletitle{A study of snippet length and informativeness: Behaviour, performance and user experience}. In \bibinfo{booktitle}{\emph{Proceedings of the 40th International ACM SIGIR Conference on Research and Development in Information Retrieval}}. \bibinfo{pages}{135--144}.
\newblock


\bibitem[Oliveira and Teixeira~Lopes(2023)]%
        {oliveira2023evolution}
\bibfield{author}{\bibinfo{person}{Bruno Oliveira} {and} \bibinfo{person}{Carla Teixeira~Lopes}.} \bibinfo{year}{2023}\natexlab{}.
\newblock \showarticletitle{The Evolution of Web Search User Interfaces-An Archaeological Analysis of Google Search Engine Result Pages}. In \bibinfo{booktitle}{\emph{Proceedings of the 2023 Conference on Human Information Interaction and Retrieval}}. \bibinfo{pages}{55--68}.
\newblock


\bibitem[Pattanittum et~al\mbox{.}(2021)]%
        {pattanittum2021roselle}
\bibfield{author}{\bibinfo{person}{Porjai Pattanittum}, \bibinfo{person}{Chetta Ngamjarus}, \bibinfo{person}{Fonthip Buttramee}, {and} \bibinfo{person}{Charoonsak Somboonporn}.} \bibinfo{year}{2021}\natexlab{}.
\newblock \showarticletitle{Roselle for hypertension in adults}.
\newblock \bibinfo{journal}{\emph{Cochrane Database of Systematic Reviews}} \bibinfo{number}{11} (\bibinfo{year}{2021}).
\newblock


\bibitem[Pogacar et~al\mbox{.}(2017)]%
        {pogacar2017positive}
\bibfield{author}{\bibinfo{person}{Frances~A Pogacar}, \bibinfo{person}{Amira Ghenai}, \bibinfo{person}{Mark~D Smucker}, {and} \bibinfo{person}{Charles~LA Clarke}.} \bibinfo{year}{2017}\natexlab{}.
\newblock \showarticletitle{The positive and negative influence of search results on people's decisions about the efficacy of medical treatments}. In \bibinfo{booktitle}{\emph{Proceedings of the ACM SIGIR International Conference on Theory of Information Retrieval}}. \bibinfo{pages}{209--216}.
\newblock


\bibitem[Robertson et~al\mbox{.}(1995)]%
        {robertson1995okapi}
\bibfield{author}{\bibinfo{person}{Stephen~E Robertson}, \bibinfo{person}{Steve Walker}, \bibinfo{person}{Susan Jones}, \bibinfo{person}{Micheline~M Hancock-Beaulieu}, \bibinfo{person}{Mike Gatford}, {et~al\mbox{.}}} \bibinfo{year}{1995}\natexlab{}.
\newblock \showarticletitle{Okapi at TREC-3}.
\newblock \bibinfo{journal}{\emph{Nist Special Publication Sp}}  \bibinfo{volume}{109} (\bibinfo{year}{1995}), \bibinfo{pages}{109}.
\newblock


\bibitem[Rose et~al\mbox{.}(2007)]%
        {rose2007summary}
\bibfield{author}{\bibinfo{person}{Daniel~E Rose}, \bibinfo{person}{David Orr}, {and} \bibinfo{person}{Raj Gopal~Prasad Kantamneni}.} \bibinfo{year}{2007}\natexlab{}.
\newblock \showarticletitle{Summary attributes and perceived search quality}. In \bibinfo{booktitle}{\emph{Proceedings of the 16th international conference on World Wide Web}}. \bibinfo{pages}{1201--1202}.
\newblock


\bibitem[Stamou and Efthimiadis(2010)]%
        {stamou2010interpreting}
\bibfield{author}{\bibinfo{person}{Sofia Stamou} {and} \bibinfo{person}{Efthimis~N Efthimiadis}.} \bibinfo{year}{2010}\natexlab{}.
\newblock \showarticletitle{Interpreting user inactivity on search results}. In \bibinfo{booktitle}{\emph{European Conference on Information Retrieval}}. Springer, \bibinfo{pages}{100--113}.
\newblock


\bibitem[Stede and Schneider(2018)]%
        {stede2018argumentation}
\bibfield{author}{\bibinfo{person}{Manfred Stede} {and} \bibinfo{person}{Jodi Schneider}.} \bibinfo{year}{2018}\natexlab{}.
\newblock \showarticletitle{Argumentation mining}.
\newblock \bibinfo{journal}{\emph{Synthesis Lectures on Human Language Technologies}} \bibinfo{volume}{11}, \bibinfo{number}{2} (\bibinfo{year}{2018}), \bibinfo{pages}{1--191}.
\newblock


\bibitem[Tombros and Sanderson(1998)]%
        {tombros1998advantages}
\bibfield{author}{\bibinfo{person}{Anastasios Tombros} {and} \bibinfo{person}{Mark Sanderson}.} \bibinfo{year}{1998}\natexlab{}.
\newblock \showarticletitle{Advantages of query biased summaries in information retrieval}. In \bibinfo{booktitle}{\emph{Proceedings of the 21st annual international ACM SIGIR conference on Research and development in information retrieval}}. \bibinfo{pages}{2--10}.
\newblock


\bibitem[Wason(1960)]%
        {wason1960failure}
\bibfield{author}{\bibinfo{person}{Peter~C Wason}.} \bibinfo{year}{1960}\natexlab{}.
\newblock \showarticletitle{On the failure to eliminate hypotheses in a conceptual task}.
\newblock \bibinfo{journal}{\emph{Quarterly journal of experimental psychology}} \bibinfo{volume}{12}, \bibinfo{number}{3} (\bibinfo{year}{1960}), \bibinfo{pages}{129--140}.
\newblock


\bibitem[White(2013)]%
        {white2013beliefs}
\bibfield{author}{\bibinfo{person}{Ryen White}.} \bibinfo{year}{2013}\natexlab{}.
\newblock \showarticletitle{Beliefs and biases in web search}. In \bibinfo{booktitle}{\emph{Proceedings of the 36th international ACM SIGIR conference on Research and development in information retrieval}}. \bibinfo{pages}{3--12}.
\newblock


\bibitem[White and Hassan(2014)]%
        {white2014content}
\bibfield{author}{\bibinfo{person}{Ryen~W White} {and} \bibinfo{person}{Ahmed Hassan}.} \bibinfo{year}{2014}\natexlab{}.
\newblock \showarticletitle{Content bias in online health search}.
\newblock \bibinfo{journal}{\emph{ACM Transactions on the Web (TWEB)}} \bibinfo{volume}{8}, \bibinfo{number}{4} (\bibinfo{year}{2014}), \bibinfo{pages}{1--33}.
\newblock


\bibitem[White et~al\mbox{.}(2003)]%
        {white2003task}
\bibfield{author}{\bibinfo{person}{Ryen~W White}, \bibinfo{person}{Joemon~M Jose}, {and} \bibinfo{person}{Ian Ruthven}.} \bibinfo{year}{2003}\natexlab{}.
\newblock \showarticletitle{A task-oriented study on the influencing effects of query-biased summarisation in web searching}.
\newblock \bibinfo{journal}{\emph{Information Processing \& Management}} \bibinfo{volume}{39}, \bibinfo{number}{5} (\bibinfo{year}{2003}), \bibinfo{pages}{707--733}.
\newblock


\bibitem[Wu et~al\mbox{.}(2020)]%
        {wu2020credibility}
\bibfield{author}{\bibinfo{person}{Dan Wu}, \bibinfo{person}{Jing Dong}, \bibinfo{person}{Li Shi}, \bibinfo{person}{Chunxiang Liu}, {and} \bibinfo{person}{Jiangyun Ding}.} \bibinfo{year}{2020}\natexlab{}.
\newblock \showarticletitle{Credibility assessment of good abandonment results in mobile search}.
\newblock \bibinfo{journal}{\emph{Information Processing \& Management}} \bibinfo{volume}{57}, \bibinfo{number}{6} (\bibinfo{year}{2020}), \bibinfo{pages}{102350}.
\newblock


\end{thebibliography}
